\documentstyle[12pt,psfig]{article}
\begin{document}

\begin{center}
{\large \bf Polaron Variational Methods In The Particle
Representation Of Field Theory :  II. Numerical Results For The
Propagator}

\vspace{1.5cm}
R.~Rosenfelder $^{\star}$ and A.~W.~Schreiber $^{\star \> \> \dagger}$

\vspace{1cm}
$^{\star}$ Paul Scherrer Institute, CH-5232 Villigen PSI, Switzerland

$^{\dagger}$ TRIUMF, 4004 Wesbrook Mall, Vancouver, B.C.,
Canada V6T 2A3

\end{center}

\vspace{2cm}
\begin{abstract}
\noindent
For the scalar Wick-Cutkosky model in the particle representation
we perform a similar variational calculation for the 2-point function
as was done by Feynman for the polaron problem. We employ a quadratic
nonlocal trial action with a retardation function for which
several ans\"atze are used. The variational
parameters are determined by minimizing  the variational function
and in the most general case the nonlinear variational
equations are solved numerically. We obtain
the residue at the pole, study analytically and numerically
the instability of the model at larger coupling constants and
calculate the width of the dressed particle.

\end{abstract}

\vspace{1.5cm}


\newpage

\section{Introduction}

\noindent
The need of nonperturbative methods in in quantum physics is obvious
considering the many problems where strong coupling and/or binding
effects
render perturbation theory inadequate. In nonrelativistic
many-body physics  variational methods are widely used
under these circumstances while this is {\it not} the case in
 relativistic field theory. However, Feynman's successful treatment
of the polaron problem \cite{Fey} shows that variational
methods may also be used for a nonrelativistic field theory
provided that the fast degrees of freedom can be integrated out and
their effect properly taken into account in the trial action.
In a previous paper \cite{RoSchr} (henceforth referred to as (I) )
we have extended the polaron variational method to the simplest
scalar field theory
which describes heavy particles (``nucleons'') interacting by the
exchange of light particles (``mesons''). In euclidean space-time the
Lagrangian of the Wick-Cutkosky model is given by
\begin{equation}
{\cal L} = \frac{1}{2} \left ( \partial_{\mu} \Phi \right )^2 +
\frac{1}{2}
M_0^2 \Phi^2 + \frac{1}{2} \left ( \partial_{\mu} \varphi \right )^2
 + \frac{1}{2}m^2 \varphi^2 - g \Phi^2 \varphi \> .
\end{equation}
Here $M_0$ is the bare mass of the nucleon, $m$ is the mass of the
meson and $g$ is the (dimensionfull) coupling constant of the Yukawa
interaction
between them. In the quenched approximation the meson field can be
integrated out and one ends up with an effective action for the
nucleons only
\begin{equation}
S_{\rm eff} \> [x(\tau)] = \int_0^{\beta} d \tau  \> \frac{1}{2}
\dot x^2
- \frac{g^2}{2} \int_0^{\beta} d \tau_1 \> \int_0^{\beta} d \tau_2 \>
\int \frac{d^4 q}{(2 \pi)^4} \> \frac{1}{q^2 + m^2} \>
e^{ \> i q \cdot ( \>x(\tau_1) - x(\tau_2) \> ) }  \> .
\label{eff action}
\end{equation}
Note that this is formulated in terms of trajectories $ x(\tau) $
of the
heavy particle (``particle representation'') which are parametrized by
the proper time $ \tau $ and obey the boundary conditions
$ x(0) = 0 $ and $ x(\beta) = x$. To obtain the 2-point function one
has to perform the path integral over all trajectories and
to integrate over $ \beta $ from zero to infinity with a certain
weight.
It is, of course, impossible to perform this path integral exactly
and, again following Feynman, we have approximated it variationally
by a retarded quadratic two-time action.
In (I) we have proposed various parameterizations for the
retardation function which enters this trial action and
derived variational equations for the most general case when its
form was left free.

The purpose of the present paper is to investigate numerically these
parametrizations as well as to solve the variational equations. This
 fixes
the variational parameters which will be used to calculate physical
observables in forthcoming applications. One quantity which we
evaluate
in the present paper is the residue on the pole of the propagator.
Another one is related to the well-known instability \cite{Baym}
of the Wick-Cutkosky
model : although the effective action (\ref{eff action})
is very similar to the one in the
polaron model the ground state of the system is only metastable.
This does not show up in any order of perturbation theory but, as
we have demonstrated in (I),
the variational approach knows about it. Indeed
an approximate solution of the variational equations
has revealed that there are no real solutions beyond a certain
critical coupling.
In the present paper we will find exact numerical values for
this critical coupling and calulate
the width of the unstable particle for couplings beyond it.

This paper is organized as follows: The essential points of the
polaron variational approach are collected
in Section \ref{sec: polaron},
while Section \ref{sec: numerical}  is devoted to the numerical
methods and
results. In Section \ref{sec: width} we investigate the instability
of the Wick-Cutkosky model in our variational method and
determine analytically and numerically the width of the dressed
particle. The variational principle can also be applied away from
the pole, which is explored in Section \ref{sec: var 2point off pole}
 and used to calculate the residue
at the nucleon pole. The main results of this work are
summarized in the last Section.

\section{Polaron Variational Approach}
\label{sec: polaron}

\noindent
Following Feynman's treatment of the polaron problem we have
performed in (I) a variational calculation of the 2-point function
with the quadratic trial action
\begin{equation}
S_t[x] \> = \> \int_0^{\beta} d\tau \> \frac{1}{2} \dot x^2 \> + \>
\int_0^{\beta} d\tau_1 \> \int_0^{\tau_1} d\tau_2 \>
f ( \tau_1 - \tau_2 ) \> \left [ \> x(\tau_1) \> - \> x(\tau_2) \>
 \right ]^2 \> .
\label{general x Feynman action}
\end{equation}
Here $f ( \tau_1 - \tau_2 ) $ is an undetermined `retardation
function' which takes into account the time lapse occurring when
mesons are emitted and absorbed on the nucleon. In actual
calculations we rather have used the Fourier space form
\begin{equation}
S_t = \> \sum_{k=0}^{\infty} A_k \> b_k^2 \> \>,
\label{Feynman Fourier action}
\end{equation}
where the $b_k$ are the Fourier components of the path $x(\tau)$ and
the Fourier
coefficients $A_k$ are considered as variational parameters.
The variational treatment is based on the decomposition of the action
$ S_{\rm eff} $
into $ S_{\rm eff} =  S_t \> + \Delta S $ and on Jensen's inequality
\begin{equation}
 < e^{ - \Delta S } >  \> \> \ge \> \> e^{ - < \Delta S > } \> .
\label{jensen}
\end{equation}
Near $ p^2 = - M_{\rm phys}^2 $ the 2-point function should behave
like
\begin{equation}
G_2(p^2) \longrightarrow
\>\> \frac{Z}{p^2 + M_{\rm phys}^2} \> .
\label{pole of 2point(p)}
\end{equation}
where $ 0 < Z < 1 $ is the residue. As was shown in (I) this requires
the proper time $\beta$ to tend to infinity. One then obtains the
following inequality
\begin{equation}
M_{\rm phys}^2 \le \> \frac{M_1^2}{2 \lambda} \> + \>
\frac{\lambda}{2} \> M_{\rm phys}^2 \> + \> \frac{1}{\lambda}
\> \left (\Omega \> + \> V \right )\>.
\label{var inequality for Mphys}
\end{equation}
where
\begin{equation}
M_1^2 = M_0^2 \> - \> \frac{g^2}{4 \pi^2} \> \ln \frac{\Lambda^2}{m^2}
\label{finite mass}
\end{equation}
is a finite mass into which the divergence of the self-energy
has been absorbed and $\lambda$ a variational parameter. For
$\beta \to \infty$ all discrete sums over Fourier modes $A_k$
turn into integrals over the `profile function'
$ A( E = k \pi /\beta ) $ and one finds
\begin{equation}
\Omega \> = \> \frac{2}{\pi} \> \int_0^{\infty} dE \>
\left [ \> \ln A(E) \> + \> \frac{1}{A(E)} \> - \> 1 \> \right ] \> ,
\label{Omega by A(E)}
\end{equation}
as well as
\begin{equation}
V  \> = \> - \> \frac{g^2}{8 \pi^2} \> \int_0^{\infty} d\sigma\>
\int_0^1 du \>
\Biggl [ \> \frac{1}{\mu^2(\sigma)} e \> \left ( m \mu(\sigma),
\frac{\lambda M_{\rm phys} \sigma}{ \mu(\sigma)}, u \right)
\> - \> \frac{1}{\sigma} \> e \> ( m \sqrt{\sigma},0,u)
\> \Biggr ] \> .
\label{pot}
\end{equation}
Here we use the abbreviations
\begin{equation}
 e(s,t,u) =
\exp \left ( -  \> \frac{s^2}{2} \> \frac{1-u}{u} \> - \>
\frac{t^2}{2}\> u \> \right )
\end{equation}
and
\begin{equation}
\mu^2(\sigma) \> =
\>  \frac{4}{\pi} \int_0^{\infty} dE \> \frac{1}{A(E)} \>
\frac{\sin^2 (E \sigma/ 2)}{E^2} \>.
\label{amu2(sigma)}
\end{equation}
Because  $ \mu^2(\sigma) $ behaves like $ \sigma $ and $ \sigma/A_0 $
for small and large $ \sigma $, respectively,
we have called it a `pseudotime'. Note that in Eq. (\ref{pot}) the
particular renormalization point $\mu_0 = 0$ has been used to
regularize
the small-$\sigma$ behaviour of the integrand. As we have shown in (I)
the total result is, of course, independent of $\mu_0$.

\noindent
The profile function $A(E)$ is linked to the retardation function
$f(\sigma)$ by
\begin{equation}
A( E ) =\>  1 \> + \> \frac{8}{E^2} \> \int_0^{\infty}
d\sigma \> f(\sigma) \>  \sin^2 \frac{E \sigma}{ 2 } \> .
\label{A(E)}
\end{equation}
In (I) we have studied the following parametrizations

\vspace{0.5cm}

\noindent
{\it`Feynman' parametrization:}

\begin{equation}
f_F(\sigma) \> = \> w \> \frac{v^2 - w^2}{ w} \> e^{-w \sigma} \> ,
\label{Feynman retard func}
\end{equation}
which leads to
\begin{equation}
A_F ( E ) \> = \> \frac{v^2 \> + \> E^2}{w^2 \> + \> E^2}\;\;\;.
\label{Feynman A(E)}
\end{equation}

\noindent
{\it`Improved' parametrization:}

\begin{equation}
f_I(\sigma) \> = \> \frac{v^2 - w^2}{2 w} \frac{1}{\sigma^2} \>
e^{ - w \sigma} \> ,
\label{improved retard func}
\end{equation}
which entails
\begin{equation}
A_I( E ) = \> 1 + 2 \> \frac{ v^2 - w^2}{ w E} \> \left [ \arctan
\frac{E}{w} -\> \frac{w}{2 E} \ln \left ( 1 + \frac{E^2}{w^2}
\right ) \> \right ] \> .
\label{improved A(E)}
\end{equation}

\noindent
In both cases $ v, w $ are variational parameters whose values have to
be determined by minimizing Eq. (\ref{var inequality for Mphys}).

\vspace{0.5cm}
\noindent
As well as the above parametrizations, it was possible not to
impose any specific form for the retardation function
but to determine it from varying Eq. (\ref{var inequality for Mphys})
 with respect
to $\lambda$ and $A(E)$. This gave the following relations
\begin{equation}
\frac{1}{\lambda} = \> 1 + \frac{g^2}{8 \pi^2} \> \int_0^{\infty}
d\sigma\> \frac{\sigma^2}{\mu^4(\sigma)} \>
\int_0^1 du \> u \> e \> \left ( m \mu(\sigma),
\frac{\lambda M_{\rm phys} \sigma}{ \mu(\sigma)}, u \right)
\label{var eq for lambda}
\end{equation}
\begin{eqnarray}
A(E) \> = \> 1 + \frac{g^2}{4 \pi^2} \frac{1}{E^2}
\int_0^{\infty} d\sigma \>
\frac{\sin^2 (E \sigma /2)}{\mu^4(\sigma)} \> \int_0^1 &du& \left [ 1
+ \frac{m^2}{2}
\mu^2(\sigma) \frac{1-u}{u} -\frac{\lambda^2 M^2_{\rm phys} \sigma^2}
{2 \mu^2(\sigma)} u \right ] \nonumber \\
&& \hspace{1 cm} \cdot \> e \> \left ( m \mu(\sigma),
\frac{\lambda M_{\rm phys} \sigma}{ \mu(\sigma)}, u \right) \> .
\label{var eq for A(E)}
\end{eqnarray}
Together with Eq. (\ref{amu2(sigma)}) they constitute a system
of coupled variational equations which have to be solved.
Assuming $\mu^2(\sigma) \simeq \sigma$ and $ m \simeq 0$ we have
found in (I) an approximate
solution  which
had the same form of the retardation function as the `improved'
parametrization and
exhibited the instability of the system beyond a critical coupling
constant. In the general case we can read off the variational
retardation function from Eq. (\ref{var eq for A(E)})
\begin{equation}
f_{\rm var} (\sigma) = \frac{g^2}{32 \pi^2} \> \frac{1}
{\mu^4(\sigma)} \> \int_0^1 du \left [ 1 + \frac{m^2}{2}
\mu^2(\sigma) \frac{1-u}{u} -\frac{\lambda^2 M^2_{\rm phys} \sigma^2}
{2 \mu^2(\sigma)} u \right ] \> e \> \left ( m \mu(\sigma),
\frac{\lambda M_{\rm phys} \sigma}{ \mu(\sigma)}, u \right) \> .
\label{var retardation function}
\end{equation}
Obviously it
has the same $1/\sigma^2$-behaviour for small
relative times as the `improved' parametrization
(\ref{improved retard func}). Finally we mention that by means of the
variational Eq. (\ref{var eq for A(E)}),
one can find the following expression for the the `kinetic term'
$\Omega$ defined in Eq. (\ref{Omega by A(E)})
\begin{eqnarray}
\Omega_{\rm var} = \frac{g^2}{8 \pi^2}
\int_0^{\infty} &d\sigma& \> \int_0^1 du \> \left [ 1
+ \frac{m^2}{2}
\mu^2(\sigma) \frac{1-u}{u} -\frac{\lambda^2 M^2_{\rm phys} \sigma^2}
{2 \mu^2(\sigma)} u \right ] \nonumber \\
&\cdot& e \> \left ( m \mu(\sigma),
\frac{\lambda M_{\rm phys} \sigma}{ \mu(\sigma)}, u \right) \>
\frac{\partial}{\partial \sigma} \left (   \frac{\sigma}
{\mu^2(\sigma)} \right ) \> .
\label{Omega var expressed by g^2}
\end{eqnarray}
This is demonstrated in the Appendix  and will be used in Chapter
\ref{sec: width}.
That the kinetic term $\Omega$  can be combined with the
`potential term' $V$
is a consequence of the virial theorem for a two-time action
\cite{AlRo} which the variational approximation fulfills.

\section{Numerical Results}
\label{sec: numerical}

\noindent
In this Section we will compare numerically the various
parametrizations for the retardation function.
Because we are primarily interested in an
eventual application in pion-nucleon physics, we have chosen the
masses and coupling constants appropriately.  Of course, the model
does not really give a
realistic description of the pion-nucleon interaction as spin- and
isospin degrees of freedom as well as chiral symmetry are missing.

In short, we use
\begin{eqnarray}
m \> = \> 140 \>\>\> {\rm MeV}
\label{pion mass} \\
M_{\rm phys} \> = \> 939 \>\>\> {\rm MeV}
\label{nucleon mass}
\end{eqnarray}
and the
results are presented as function of the dimensionless coupling
constant
\begin{equation}
\alpha \> = \> \frac{g^2}{4 \pi} \frac{1}{M^2_{\rm phys}} \> \> \> .
\label{alpha}
\end{equation}
The relevant quantity for the physical situation is the
strength of the Yukawa potential between two nucleons
due to one-pion exchange \cite{BjDr}, which is approximately given by
(depending on the spin-isospin channel)
\begin{equation}
f^2 = \frac{g'^2}{4 \pi} \> \left ( \frac{m}{2 M_{\rm phys}}
\right )^2 \> \cong \> 0.08\;\;\;,
\label{pion-nucleon coupling}
\end{equation}
where $ \> g'^2/4 \pi \cong 14 \> $ is the pion-nucleon coupling.
In the Wick-Cutkosky scalar model the corresponding strength is just
the dimensionless coupling constant $ \alpha$ that
we have defined in Eq. (\ref{alpha}).
It should also be remembered
that a Yukawa potential only supports a bound state~\cite{ErWe}  if
\begin{equation}
\alpha \> > \> 1.680 \> \frac{m}{M} \> = 0.2505 \> .
\label{critical binding for Yukawa}
\end{equation}

\noindent
We have minimized (cf. Eq. (\ref{var inequality for Mphys}))
\begin{equation}
 - \> M_1^2 \> \le \> (\lambda^2 - 2 \lambda) \> M^2_{\rm phys}
\> + \> 2 \left ( \Omega \> + \> V \right )
\label{minimization}
\end{equation}
with the `Feynman' ansatz (\ref{Feynman A(E)}) and the `improved'
ansatz (\ref{improved A(E)}).  This minimization was performed
numerically with respect to the
parameters $\lambda, v, w$ by using the standard CERN program MINUIT.
The numerical integrations were done with typically $2 \times 72$
Gauss-Legendre points after
mapping the infinite-range integrals to finite range. For the
`improved' retardation function we had to calculate $\mu^2(\sigma)$
and $\Omega$ numerically. Tables~\ref{table: var Feyn}
and~\ref{table: var improved} give the
results of these calculations. We also include the value of $M_1$
although it doesn't have a physical meaning: finite terms (which,
for example, arise when a different renormalization point is chosen)
can be either grouped with $M_1$ or with $V$. However, from the
variational inequality (\ref{minimization}) we see that
$M_1$ is a measure of the quality of the variational approximation:
the larger $M_1$ the better the approximation.

\begin{table}
\begin{center}
\begin{tabular}{|c|c|c|c|c|c|} \hline
 ~~$ \alpha $~~ & $ \sqrt v $ \ [MeV] &$  \sqrt w $ \ [MeV] &
$ \lambda  $ & $ M_1 $ \ [MeV]& $   A(0)  $ \\ \hline
 0.1     & 1850    & 1845       & ~0.97300~  & 890.23 & ~1.0120~ \\
 0.2     & 1805    & 1794       & ~0.94400~  & 839.73 & ~1.0257~ \\
 0.3     & 1756    & 1739       & ~0.91250~  & 787.29 & ~1.0417~ \\
 0.4     & 1702    & 1678       & ~0.87773~  & 732.69 & ~1.0606~ \\
 0.5     & 1641    & 1608       & ~0.83843~  & 675.70 & ~1.0838~ \\
 0.6     & 1569    & 1527       & ~0.79223~  & 616.09 & ~1.1142~ \\
 0.7     & 1477    & 1424       & ~0.73355~  & 553.93 & ~1.1582~ \\
 0.8     & 1325    & 1254       & ~0.63714~  & 490.60 & ~1.2485~ \\
\hline
\end{tabular}
\end{center}
\caption{Variational calculation for the nucleon self-energy in
the Wick-Cutkosky model using the `Feynman' parametrization
(\protect\ref{Feynman A(E)}) for the profile function.
The parameters $v,w$ obtained
from minimizing Eq. (\protect\ref{minimization}) are given as
well as $\lambda$  and the intermediate renormalized mass $M_1$
(see Eq. (\protect\ref{finite mass})). The last column lists
$ A(0) = v^2/w^2$.}
\label{table: var Feyn}
\end{table}

\begin{table}
\begin{center}
\begin{tabular}{|c|c|c|c|c|c|} \hline
 ~~$\alpha$ ~~ &$\sqrt v$\ [Mev]&$\sqrt w$\ [MeV]
&$\lambda$&$M_1$ \ [MeV]  & $ A(0) $ \\ \hline
 0.1     & 677.2  & 674.6      & ~0.97297~  & 890.25 & ~1.0158~ \\
 0.2     & 661.4  & 656.0      & ~0.94390~  & 839.78 & ~1.0338~ \\
 0.3     & 640.8  & 632.3      & ~0.91223~  & 787.43 & ~1.0548~ \\
 0.4     & 613.7  & 601.9      & ~0.87715~  & 732.97 & ~1.0808~ \\
 0.5     & 596.7  & 581.2      & ~0.83741~  & 676.20 & ~1.1109~ \\
 0.6     & 570.4  & 550.7      & ~0.79040~  & 616.97 & ~1.1514~ \\
 0.7     & 534.3  & 509.2      & ~0.72996~  & 555.45 & ~1.2118~ \\
 0.8     & 468.2  & 434.5      & ~0.62429~  & 493.44 & ~1.3482~ \\
\hline
\end{tabular}
\end{center}
\caption{
Same as in Table \protect\ref{table: var Feyn} but
using the `improved' parameterization
(\protect\ref{improved A(E)}) for the profile function. }
\label{table: var improved}
\end{table}

Although the value of the parameters $v$ and $w$ are rather different
for the Feynman and the `improved' parametrization, the parameter
$\lambda$ and the value of the profile function at $E = 0$ are very
close. This is also reasonable when we study the behaviour of these
quantities under
a {\it reparametrization} of the particle path: it can be shown that
a rescaling of the proper time $\beta \to \beta/\kappa$
leaves the variational functional invariant if
\begin{equation}
A^{(\kappa)} \left ( \frac{E}{\kappa} \right ) \> = \>
A^{(\kappa = 1)} ( E ) \> .
\label{repar invariance}
\end{equation}
We are working in the `proper time gauge' $ \kappa = 1 $. In a
general `gauge'
$ \kappa $ the variational parameters $ v, w$ then obviously are
different ( see Eqs. (\ref{Feynman A(E)}, \ref{improved A(E)}) )
\begin{equation}
v^{(\kappa)} \> = \> \kappa \> v \> , \hspace{2cm}
w^{(\kappa)} \> = \> \kappa \> w \> ,
\end{equation}
but $A(0) = v^2 / w^2 $ and $\lambda$ are gauge-invariant.

For both parametrization no minimum of Eq. (\ref{minimization}) was
found beyond
\begin{equation}
\alpha \> > \> \alpha_c
\end{equation}
where
\begin{eqnarray}
\alpha_c \> = \> \left \{ \begin{array}{ll}
                          0.824   &\hspace{1 cm}
                                          \mbox{( `Feynman' )} \\
                          0.817   &\hspace{1 cm}
                                          \mbox{( `improved' )} \> .
                          \end{array}
                          \right .
\end{eqnarray}
This value of the critical coupling is surprisingly close
to the value $ \alpha_c \simeq \pi/4 $ which
we obtained from the approximate solution of the variational
equations in (I).
On the other hand when the parameter $\lambda$ is fixed to
$\lambda = 1$, i.e. a less general trial action for
"momentum averaging" (see (I)) is used, then
a minimum is found for {\it all} values of $\alpha$. This points to
the important role played by this parameter. Indeed, in the
approximate solution of the variational equations found in (I) the
branching of the real solutions into complex
ones is most clearly seen in the approximate solution for $\lambda$.
We can also trace the instability to the inequality
(\ref{var inequality for Mphys}) for the physical mass : a clear
minimum as a function of $\lambda$ exists only as long as
the coefficient of $1/\lambda$ ,
i.e. $M_1^2/2 + \Omega + V$ stays positive. However, with increasing
coupling $M_1$ shrinks and $V$ becomes more negative until
at the critical coupling the collapse finally occurs.

We have also solved the coupled nonlinear variational equations
(\ref{var eq for lambda}), (\ref{var eq for A(E)})
together with (\ref{amu2(sigma)}) numerically \footnote{Note that the
variational solution is also reparametrization
invariant: Eqs. (\ref{var eq for A(E)}) and (\ref{amu2(sigma)}) are
consistent with the condition (\ref{repar invariance}). } .
This was done
by the following {\em iterative} method: we first mapped variables
with infinite range to finite range~, e.g.
\begin{eqnarray}
E &=& M^2_{\rm phys} \> \tan \theta  \\
\sigma &=& \frac{1}{M^2_{\rm phys}} \> \tan \psi
\end{eqnarray}
and then discretized the integrals by the standard
Gauss-Legendre integration scheme, with typically 72 or 96 gaussian
points per integral.
The functions $A(\theta), \mu^2(\psi)$
as given by the variational equations were then tabulated at the
gaussian points
using as input the values of $\lambda, A(\theta), \mu^2(\psi) $ from
the previous iteration. We started with the perturbative values
\begin{eqnarray}
\lambda^{(0)} &=& A (\theta_i)^{(0)} = 1 \nonumber \\
\mu^2(\psi_i)^{(0)} &=& \frac{1}{M^2_{\rm phys}} \> \tan \psi_i
\end{eqnarray}
and monitored the convergence with the help of
the largest relative deviation
\begin{equation}
\Delta_n = {\rm Max} \left ( \frac{ | \lambda^{(n)} -
\lambda^{(n-1)} | }
{\lambda^{(n)}} \> , \> \frac{ |A (\theta_i)^{(n)} -
A (\theta_i)^{(n-1)} | }
{A (\theta_i)^{(n)}}\> , \> \frac{ | \mu^2(\psi_i)^{(n)} -
\mu^2(\psi_i)^{(n-1)} | }
{\mu^2(\psi_i)^{(n)}} \right )\> , \>\>\> n = 1, 2 ...
\label{max dev}
\end{equation}
Some numerical results are given in Table~\ref{table: var equations}.
Comparing with Table~\ref{table: var improved} we observe a remarkable
agreement with the values from the `improved' parametrization.
It is only near $\alpha = 0.8$ that the variational solution is
appreciably better as demonstrated by the
numerical value of $M_1$ which measures the quality of the
corresponding approximation.

\begin{table}
\begin{center}
\begin{tabular}{|c|c|c|c|} \hline
 ~~$\alpha$~~ & $\lambda$  &  $M_1$ \ [MeV]  & $ A(0)  $ \\ \hline
 0.1      & ~0.97297~    & 890.25          & ~1.0151~ \\
 0.2      & ~0.94389~    & 839.78          & ~1.0322~ \\
 0.3      & ~0.91223~    & 787.43          & ~1.0520~ \\
 0.4      & ~0.87718~    & 732.97          & ~1.0755~ \\
 0.5      & ~0.83738~    & 676.20          & ~1.1044~ \\
 0.6      & ~0.79030~    & 616.98          & ~1.1421~ \\
 0.7      & ~0.72968~    & 555.47          & ~1.1972~ \\
 0.8      & ~0.62262~    & 493.55          & ~1.3188~ \\ \hline
\end{tabular}
\end{center}
\caption{
The variational parameter $\lambda$, the renormalized mass $M_1$
and the value of the profile function
at $ E = 0 $ from the solution of the variational equations .}
\label{table: var equations}
\end{table}

\begin{figure}
\unitlength1mm
\begin{picture}(110,65)
\put(370,145){\makebox(110,65)
{\psfig{figure=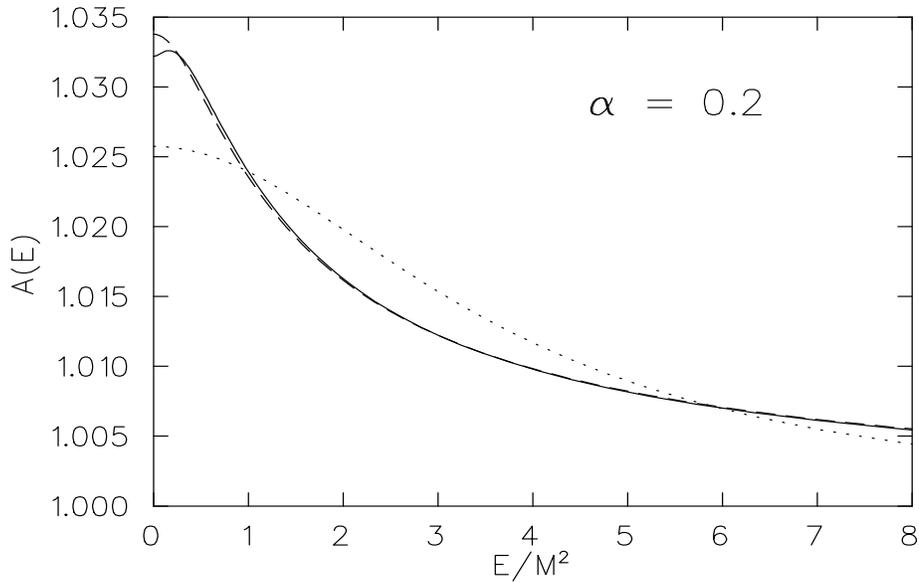,height=300mm,width=500mm}}}
\end{picture}
\caption{Profile function $A(E)$ as function of $E$ for the
`Feynman' parameterization (\protect\ref{Feynman A(E)}) (dotted line), the
`improved'
parameterization (\protect\ref{improved A(E)}) (dashed line) and the
`variational'
solution (solid line). The dimensionless coupling constant is $\alpha = 0.2$.}
\label{fig: A(E) for alpha small}
\end{figure}

This may also be seen in Figs.~\ref{fig: A(E) for alpha small},
{}~\ref{fig: mu2 for alpha small} and ~\ref{fig: alpha big}
where the different profile functions and pseudotimes
are plotted for $\alpha = 0.2$  and $\alpha = 0.8$. One can also
confirm from
the graphs that the numerical results indeed have the limits for
$\sigma, E$ either small or large which we
expect from the analytical analysis. Furthermore, it
is clear that the `improved' parametrization of the trial action
is in general extremely close to the `variational' one, while the
`Feynman' parametrization deviates much more. Finally, it is
interesting to note that the profile
function of the `variational' calculation has a small dip near $E = 0$
which is a result of the additional terms in the retardation function
(\ref{var retardation function}). These rather innocent looking
deviations will become important if an analytic continuation back
to Minkowski space is performed in which physical scattering
processes take place.

\begin{figure}
\unitlength1mm
\begin{picture}(110,65)
\put(370,145){\makebox(110,65)
{\psfig{figure=musplot2.ps,height=300mm,width=500mm}}}
\put(348,200){\makebox(40,40)
{\psfig{figure=musplot2lim.ps,height=300mm,width=500mm}}}
\end{picture}
\caption{Ratio of pseudotime $\mu^2(\sigma)$ to proper time
$\sigma$ for $\alpha = 0.2$ . The labeling of the curves is as in Fig.
\protect\ref{fig: A(E) for alpha small} .  An expanded view of
the small-$\sigma$
region is shown in the inset.}
\label{fig: mu2 for alpha small}
\end{figure}

\begin{figure}
\unitlength1mm
\begin{picture}(110,65)
\put(370,145){\makebox(110,65)
{\psfig{figure=plot8.ps,height=300mm,width=500mm}}}
\end{picture}
\caption{A(E) and $\mu^2(\sigma)$ for $\alpha = 0.8$.  The labeling of the
curves is as in Fig. \protect\ref{fig: A(E) for alpha small}.}
\label{fig: alpha big}
\end{figure}

\begin{figure}
\unitlength1mm
\begin{picture}(110,70)
\put(370,140){\makebox(110,65)
{\psfig{figure=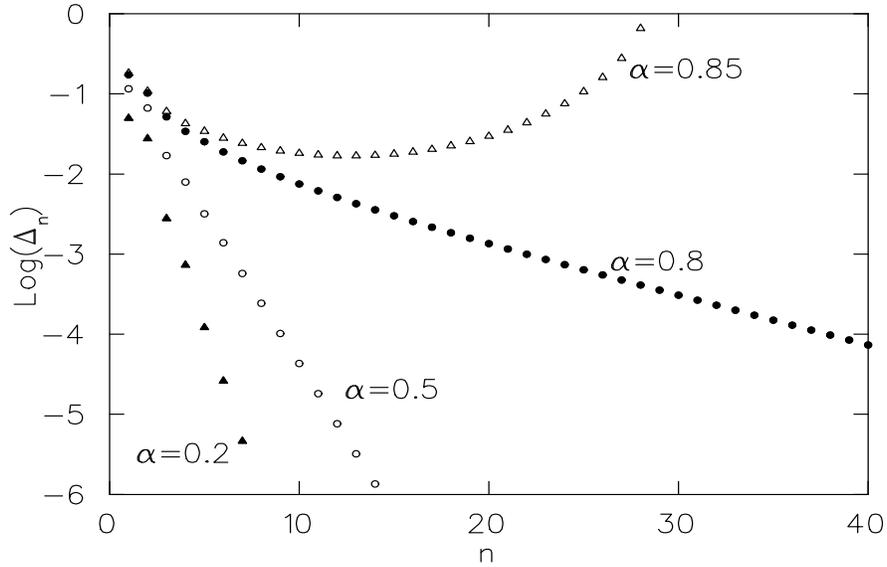,height=300mm,width=500mm}}}
\end{picture}
\caption{Convergence of the iterative solution of the
variational equations as a function of the number of iterations $n$.
The convergence measure $\Delta_n$ is defined in Eq.~(\protect\ref{max dev}).}
\label{fig: convergence}
\end{figure}

Examples for the convergence of the iterative scheme
are shown in Fig.~\ref{fig: convergence} . It is seen that for small
coupling constant we have rapid convergence which becomes slower
and slower as the critical value
\begin{equation}
\alpha_c \> = \> 0.815 \hspace{1 cm} \mbox{( `variational' )}
\end{equation}
is reached. Finally, beyond $\alpha > \alpha_c$ only a minimal
relative accuracy can be reached and the deviations increase
again with additional iterations.

\begin{figure}
\unitlength1mm
\begin{picture}(110,65)
\put(370,135){\makebox(110,65)
{\psfig{figure=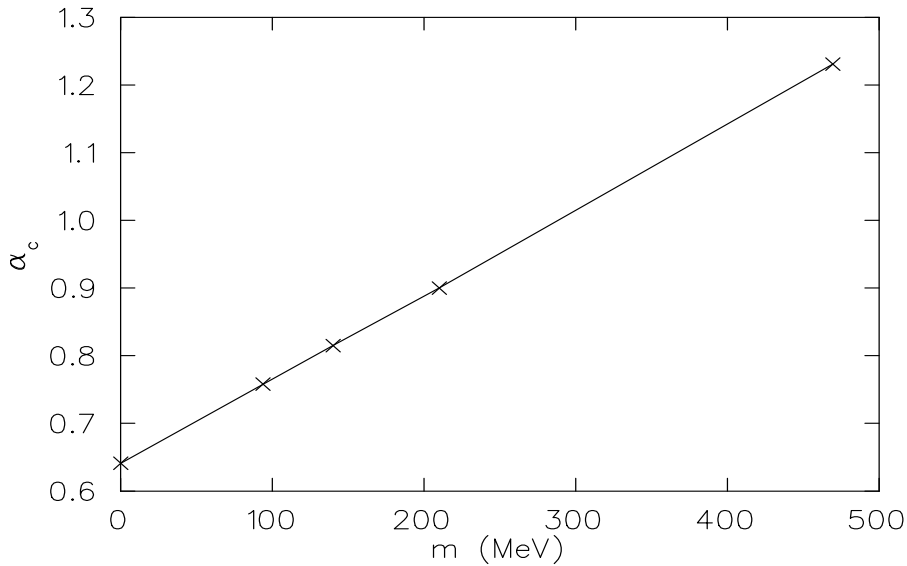,height=300mm,width=500mm}}}
\end{picture}
\caption{Critical coupling constant as a function of the meson mass $m$.
The nucleon mass is fixed at $M = 939$ MeV .  The crosses indicate the points
at which the critical coupling has been calculated, the line through them being
drawn to guide the eye.}
\label{fig: alphacrit vs m}
\end{figure}

How the critical coupling depends on the meson mass is shown in
Fig.~\ref{fig: alphacrit vs m} . It turns out that the good agreement
of the approximate value of $\alpha_c \approx \pi/4$ with the
numerical value obtained for $m = 140 $ MeV was an accidental one:
at $m = 0$
we have  $\alpha_c = 0.641$. There is also a surprisingly strong
but nearly linear $m$-dependence which we cannot reproduce from an
approximate solution of the variational equations when taking
$m \neq 0$ but still assuming $\mu^2(\sigma) \approx \sigma$.

\section{Instability and Width of the Dressed Particle}
\label{sec: width}

\noindent
In all parametrizations of the profile function $A(E)$ which we
investigated
numerically in the previous section it turned out to be impossible to
find a (real) solution of the variational equations or the variational
inequality for coupling constants above a critical value $\alpha_c$.
This is a signal of the instability of the model which is already seen
in the classical ``potential''
\begin{equation}
V^{(0)}(\Phi,\varphi) = \frac{1}{2} M_0^2 \Phi^2 \> + \> \frac{1}{2}
m^2 \varphi^2 - g \Phi^2 \> \varphi
\label{classical potential}
\end{equation}
and tells us
that the physical mass of the dressed particle becomes complex
\begin{equation}
M_{\rm phys} \> = \> M \> - \> i \> \frac{\Gamma}{2} \> .
\label{complex mass}
\end{equation}
In the following we take the real part of the nucleon mass to be
$ \> M = 939 \> $ MeV and try to determine the width $\Gamma$.

Note that in a perturbative calculation no sign of the instability
shows up : the one-loop self-energy
\begin{equation}
\Sigma(p^2) \> = \> - \frac{g^2}{4 \pi^2} \> \ln \frac{\Lambda^2}{m^2}
\> + \> \frac{g^2}{4 \pi^2} \> \int_0^1 du \>
\ln \left [ 1 + \frac{p^2}{m^2} u + \frac{M_0^2}{m^2} \frac{u}{1-u} \>
\right ] \> .
\label{perturb self energy}
\end{equation}
is perfectly well-behaved. Also the one-loop effective potential
\cite{ItZu,IIM} is not very indicative: in quenched approximation it
 is given by
\begin{equation}
V_{\rm eff}^{(1)}(\Phi,\varphi)  \> = \>
\frac{1}{2} \int \frac{d^4 p}{(2 \pi)^4} \> \ln \left [ \> 1
 \> - \>  \frac{4 g^2 \Phi^2}{p^2 + m^2} \> \frac{1}{p^2 + M_0^2 -
2 g \varphi - i \epsilon} \right ] \> .
\label{one loop Veff}
\end{equation}
A detailed analysis shows that the quantum corrections lower the
barrier which makes the ground state metastable in
$V^{(0)}(\Phi,\varphi)$
but do not remove it. In addition,
the one-loop effective potential develops an imaginary part but it
is easy to see that Im $V_{\rm eff}^{(1)}$ vanishes for
$M^2 - 2 g \Phi > 4 g^2 \phi^2 /m^2 $, i.e. it contains a
(non-analytic) step function. Therefore all proper one-loop
Green functions (which are generated by the effective action)
carry no sign of the instability.

\noindent
In contrast the variational approach for the two-point function
{\it knows} about the instability if we allow the parameter $\lambda$
in the trial action to vary.
Since the approximate solution of the variational
equation for $\lambda$ in (I) clearly showed the impossibility
to obtain a real solution beyond $ \> \alpha_c \> $,
we will first study the width
of the state using similar approximative methods before turning to the
exact numerical evaluation.

\subsection{Approximate analytical treatment}

\noindent
In order to discuss complex solutions of the variational equations
it is useful to introduce the complex quantity
\begin{equation}
\zeta \> = \> \lambda \> M_{\rm phys}
\label{def xi}
\end{equation}
and to write it in the form
\begin{equation}
\zeta \> = \> \zeta_0 \> e^{- i \chi} \> .
\label{def xi0 and chi}
\end{equation}
It is a phase $ \chi \neq 0 $ which will lead to the complex pole
(\ref{complex mass}) of the two-point function.
With the same approximation $ m = 0$ and
$ \mu^2(\sigma) \> \approx \> \sigma$ which was used before
we now obtain
\begin{equation}
\frac{1}{\lambda} \> = \> 1 \> + \> \frac{\alpha}{\pi} \>
\frac{M^2}{\zeta^2} \> .
\label{approx eq lambda xi}
\end{equation}
Here the dimensionless coupling constant is defined in terms of the
real part of the physical mass
\begin{equation}
\alpha \equiv \> \frac{g^2}{4 \pi^2 M^2} \> .
\label{def alpha for complex mass}
\end{equation}
Due to Eq. (\ref{Omega var expressed by g^2}) the kinetic term
vanishes under the same approximation
\begin{equation}
\Omega_{\rm var} \> \approx \> 0 \> .
\label{Omega approx}
\end{equation}
and the potential term becomes
\begin{eqnarray}
V &\approx& - \frac{g^2}{8 \pi^2} \int_0^1 du \> \int_0^{\infty}
d\sigma
\> \frac{1}{\sigma} \> \left [ \> \exp \left( - \frac{m^2}{2}
\frac{1-u}{u} \sigma - \frac{\zeta^2}{2} u \sigma \right ) -
\exp \left( - \frac{m^2}{2} \frac{1-u}{u} \sigma \right ) \> \right ]
\nonumber \\
&=& - \frac{g^2}{8 \pi^2} \int_0^1 du \> \ln \left [ \> 1 +
\frac{\zeta^2}{m^2} \frac{u^2}{1-u} \> \right ] \> .
\label{V approx}
\end{eqnarray}
These are rather drastic simplifications but the exact numerical
calculations show that the imaginary part of $\Omega$ is indeed
smaller (by a factor of five) than Im $V$.
Note that $V$ is not infrared stable, i.e. it diverges if the
meson mass $m$ is set to zero.
With the above approximations the stationarity equation
(\ref{var inequality for Mphys})
then reads
\begin{equation}
M_1^2 \> = \> \left ( \frac{2}{ \lambda} - 1 \right ) \> \zeta^2 +
 \frac{\alpha}{ \pi} M^2 \> \int_0^1 du \> \ln
\left [ 1 + \frac{\zeta^2}{m^2} \frac{u^2}{1-u} \right ] \> .
\end{equation}
Using Eq. (\ref{approx eq lambda xi}) this is equivalent to
\begin{equation}
\zeta^2 = M_1^2 - \frac{2 \alpha}{\pi} M^2 +
 \frac{\alpha}{ \pi} M^2 \> \int_0^1 du \> \ln
\left [ 1 + \frac{\zeta^2}{m^2} \frac{u^2}{1-u} \right ] \> .
\label{approx eq for xi}
\end{equation}
If we take the {\it imaginary} part of this equation it is possible
to set $m = 0$ and we obtain
\begin{equation}
\zeta_0^2 \> \sin 2 \chi = \frac{2 \alpha}{\pi} M^2 \> \chi \>.
\label{eq xi0 chi}
\end{equation}
How do we determine the width of the unstable state ? We take the
defining equation (\ref{def xi})
for $\> \zeta \> $, eliminate $\lambda$ by means of
Eq. (\ref{approx eq lambda xi}) and use Eq. (\ref{complex mass}).
This gives
\begin{equation}
M \> - i \> \frac{\Gamma}{2} \> = \> \zeta \> + \> \frac{\alpha}{\pi}
\frac{M^2}{\zeta} \>.
\end{equation}
The real and imaginary parts of this equation allow us to express
$\zeta_0$ and the width as a function of the
phase $\chi$. A simple calculation gives
\begin{equation}
\zeta_0 = \frac{M}{2 \cos \chi} \> \left [ \> 1 +
\sqrt{1 - \frac{4 \alpha}{\pi} \cos^2 \chi} \> \> \right ]
\label{xi0 as function of chi}
\end{equation}
and the width is
\begin{equation}
\Gamma \> = \> 2 M \tan \chi \> \sqrt{1 - \frac{4 \alpha}{\pi}
\cos^2 \chi} \>.
\label{Gamma expressed by chi and alpha}
\end{equation}
We have chosen the root which results in a positive width for
$ \> 0 \le \chi \le \pi/2 \> $.
Finally, substituting Eq. (\ref{xi0 as function of chi})
into Eq. (\ref{eq xi0 chi}) gives
the transcendental equation which determines the phase $\chi$.
After some algebraic transformations we obtain it in the form
\begin{equation}
\alpha = \pi \> \> \frac{2 \chi \> \sin 2 \chi}
{ (2 \chi + \sin 2 \chi)^2 \> \cos^2 \chi}
\> \equiv \> \frac{\pi}{4} \> h (\chi) \> .
\label{transc eq for chi}
\end{equation}
It is easy to see that the function $h (\chi)$ grows monotonically
 from $ h(0) = 1$ to $h(\pi/2) = \infty$. Solutions $\chi(\alpha)$
therefore only exist for
\begin{equation}
\alpha \>   > \> \alpha_c \> = \> \frac{\pi}{4} \> ,
\end{equation}
which is the same critical value of the coupling constant at which
previously the real (approximate) solutions of the variational
equations ceased to exist.
It is also easy to find solutions for the transcendental equation
(\ref{transc eq for chi}) for small $\chi$ : from $ h(\chi) = 1 +
\chi^2 + {\cal O}(\chi^4) $ we find
\begin{equation}
\chi \> \approx \> \sqrt{ \frac{\alpha - \alpha_c}{\alpha_c}}
\end{equation}
where, of course, $\alpha_c = \pi /4 $ should be used.
Since the expression (\ref{Gamma expressed by chi and alpha}) for
the width can be transformed into
\begin{equation}
\Gamma = 2 M \> \tan \chi \> \> \frac{ 2 \chi - \sin 2 \chi}
{2 \chi + \sin 2 \chi}
\label{Gamma expressed by chi}
\end{equation}
we obtain the following {\it nonanalytic} dependence of the width
on the coupling constant
\begin{equation}
\Gamma \> \approx \> \frac{2}{3} M \> \left ( \frac{\alpha - \alpha_c}
{\alpha_c}
\right )^{3/2} \> .
\label{Gamma for small chi}
\end{equation}
This should be valid near the critical coupling constant.

\subsection{Numerical results}

\begin{figure}
\unitlength1mm
\begin{picture}(110,70)
\put(370,140){\makebox(110,65)
{\psfig{figure=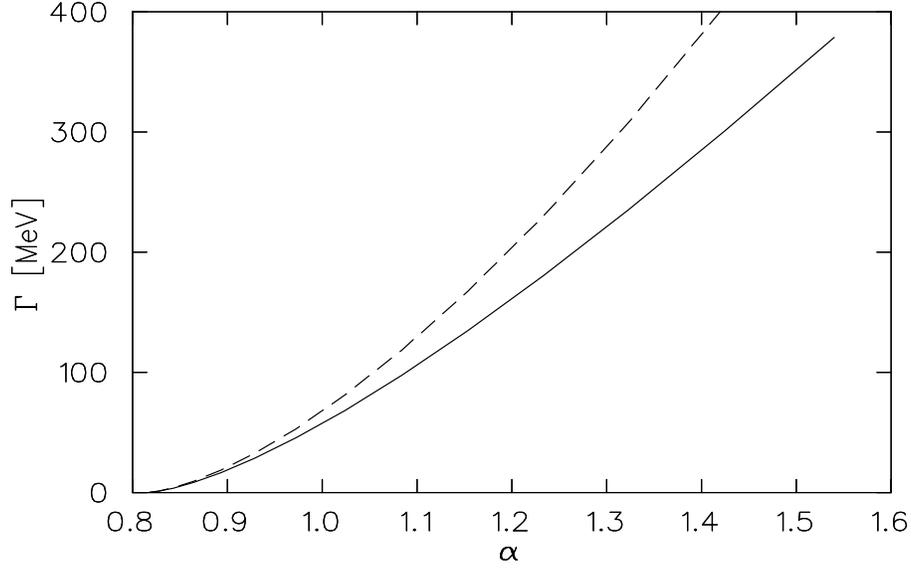,height=300mm,width=500mm}}}
\end{picture}
\caption{Width of the unstable state as a function of the coupling constant
as obtained from the solution of the complex variational equations (see Table
\protect\ref{table: width} ). The dashed line shows the approximate solution
(\protect\ref{Gamma for small chi}). }
\label{fig: width as function of alpha}
\end{figure}

\noindent
For the numerical solution of the complex variational equations
we follow the approximate analytical solution as closely as possible.
However, some of the relations used previously do not hold exactly.
For example, the quantity
\begin{equation}
L = \frac{\zeta^2}{2} \> \int_0^{\infty} d \sigma \> \frac{\sigma^2}
{\mu^4(\sigma)} \> \int_0^1 du \> u \>
e \left ( m \mu(\sigma), \frac{\zeta \sigma}{\mu(\sigma)}, u \right )
\label{def L}
\end{equation}
would be unity for $m = 0, \> \mu^2(\sigma) = \sigma $ but has some
complex value in the exact treatment. Similarly, $ \Omega \neq 0 \> $
and $ V $ deviate from the aproximate value (\ref{V approx}).
Without invoking the simplifying assumptions
Eq. (\ref{approx eq for xi}) changes to
\begin{equation}
\zeta^2 = M_1^2 - \frac{2 \alpha}{\pi} M^2 L \> + \>
2 \> ( \> \Omega \> + \> V \> )
\label{eq xi}
\end{equation}
Following the same steps as in the approximate treatment we obtain
\begin{equation}
\zeta_0 = \frac{M}{2 \cos \chi} \left [ 1 +
\sqrt{1 - \frac{4 \alpha}{\pi} \cos \chi \> {\rm Re} \> ( L
e^{i \chi} ) \> } \> \right ]
\label{xi0}
\end{equation}
which replaces Eq. (\ref{xi0 as function of chi}) and
\begin{equation}
\alpha = \pi \> \frac{K}{ \left ( \> {\rm Re} \>  ( L e^{i \chi} )
\> +\>  K \> \cos \chi \> \right )^2 }
\label{eq for alpha}
\end{equation}
which supersedes Eq. (\ref{transc eq for chi}). Here
\begin{equation}
K = \frac{2}{\sin 2 \chi} \>  {\rm Im} \> \left [ \>  L -
\frac{\pi}{\alpha} \frac{1}{M^2} \> ( \Omega + V ) \> \right ] \> .
\label{def K}
\end{equation}
Instead of Eq. (\ref{Gamma expressed by chi}) one can show that the
width itself has now the exact form
\begin{equation}
\Gamma = 2 M \> \frac{K \> \sin \chi \> - \> {\rm Im} \> ( L
e^{i \chi} ) } {K \> \cos \chi + {\rm Re} \> ( L e^{i \chi} ) } \> .
\label{Gamma exact}
\end{equation}
We have solved the coupled complex equations by specifying a value for
the phase $\chi$ and determining the corresponding value of the
coupling constant $ \alpha $ by means of Eq. (\ref{eq for alpha}).
Of course, this could be done only iteratively by starting with
\begin{eqnarray}
L^{(0)} &=& 1 \> , \> \> \> \>  K^{(0)} = 2 \chi / \sin 2 \chi  \> ,
\nonumber \\
\mu^{(0) \> 2} (\sigma) &=& \sigma \>, \> \> \> A^{(0)}(E) = 1 \> .
  \nonumber
\end{eqnarray}
Typically 20 -- 25 iterations were needed to get a relative accuracy
of better than $10^{-5}$.
Table~\ref{table: width}
 gives the results of our calculations. It is seen that
the width grows rapidly after the coupling constant exceeds the
critical value. In Fig.~\ref{fig: width as function of alpha}
 this is shown together with the approximate (small-$\chi$)
behaviour predicted by Eq. (\ref{Gamma for small chi}). After the
critical coupling constant
in this formula has been shifted to the precise value one observes a
satisfactory agreement with the exact result.

\begin{table}
\begin{center}
\begin{tabular}{|c|c|c|c|} \hline
 $\chi $ & $ \alpha $  &$ ~~\Gamma $\ [MeV]~~  & $  A(0) $  \\ \hline
 ~0.05~    & ~0.818~      &   0.13         & ~1.405 + 0.045 $i$~ \\
 ~0.10~    & ~0.827~      &   1.05         & ~1.396 + 0.088 $i$~ \\
 ~0.15~    & ~0.843~      &   3.54         & ~1.382 + 0.130 $i$~ \\
 ~0.20~    & ~0.865~      &   8.42         & ~1.362 + 0.169 $i$~ \\
 ~0.25~    & ~0.893~      &  16.5~         & ~1.338 + 0.205 $i$~ \\
 ~0.30~    & ~0.929~      &  28.6~         & ~1.309 + 0.236 $i$~ \\
 ~0.35~    & ~0.972~      &  45.5~         & ~1.277 + 0.263 $i$~ \\
 ~0.40~    & ~1.024~      &  68.2~         & ~1.243 + 0.285 $i$~ \\
 ~0.45~    & ~1.084~      &  97.6~         & ~1.207 + 0.301 $i$~ \\
 ~0.50~    & ~1.153~      & 134~~~         & ~1.171 + 0.313 $i$~ \\
 ~0.55~    & ~1.232~      & 180~~~         & ~1.134 + 0.319 $i$~ \\
 ~0.60~    & ~1.323~      & 235~~~         & ~1.099 + 0.320 $i$~ \\
 ~0.65~    & ~1.425~      & 301~~~         & ~1.065 + 0.316 $i$~ \\
 ~0.70~    & ~1.540~      & 379~~~         & ~1.032 + 0.307 $i$~ \\
\hline
\end{tabular}
\end{center}
\caption{
The width  $\Gamma $ of the unstable state from the complex solution of the
variational equations for $ \alpha  \> >  \> \alpha_c = 0.815 $. The width
is given
as a function of the phase $\chi$ which determines the corresponding
coupling constant $\alpha$ according to Eq.~(\protect\ref{eq for alpha}).
The complex value of the profile function at $ E = 0 $ is also listed.}
\label{table: width}
\end{table}

\begin{figure}
\unitlength1mm
\begin{picture}(110,65)
\put(370,145){\makebox(110,65)
{\psfig{figure=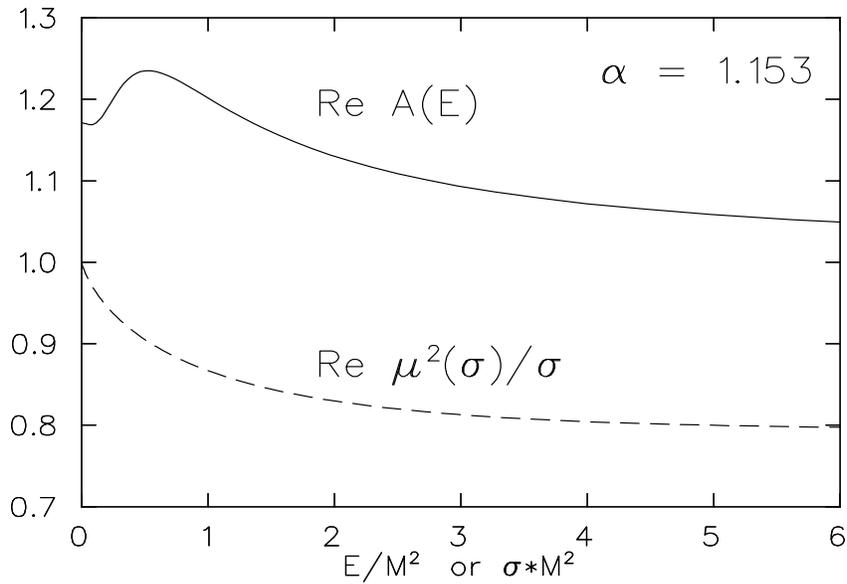,height=300mm,width=500mm}}}
\end{picture}
\caption{Real part of the profile function and of the ratio of pseudotime to
proper time for $\alpha = 1.153$. }
\label{fig: real part of A and mu2}
\end{figure}

Finally Figs.~\ref{fig: real part of A and mu2}
and~\ref{fig: im part of A and mu2} depict the complex profile
function $ A(E) $
and the
complex pseudotime $ \mu^2(\sigma) $ for $ \> \chi = 0.5 \> $, i.e.
$ \> \alpha = 1.153 \> $. Compared to the real solutions below
$ \> \alpha_c \> $ ( cf.
Figs.~\ref{fig: A(E) for alpha small}~-~\ref{fig: alpha big} ) one
does not notice any
qualitative
changes in the real part of $ A ( E ) $ as one crosses the critical
coupling.

\begin{figure}
\unitlength1mm
\begin{picture}(110,65)
\put(370,145){\makebox(110,65)
{\psfig{figure=imag.ps,height=300mm,width=500mm}}}
\end{picture}
\caption{Imaginary part of the profile function and of the ratio of pseudotime
to proper time for $\alpha = 1.153$. }
\label{fig: im part of A and mu2}
\end{figure}

\section{The Two-point Function Away from the Pole}
\label{sec: var 2point off pole}

\noindent
Up to now we only have determined the variational parameters on the
nucleon pole. However, the variational principle also applies to
$ p^2 \ne - M^2_{\rm phys}$. This forces us to consider
sub-asymptotic values of the proper time $\beta$.
We first deal with the residue at the pole which gives us the
probability to find the bare nucleon in the dressed particle.

\subsection{The residue}
\label{sec: resi}

\noindent
To calculate the residue it is most convenient to use the ``momentum
averaging'' scheme developed in (I) because in this approach there
are only a few subasymptotic terms. To be more specific the
quantity $\tilde \mu^2(\sigma,T)$ introduced in Eq. (I.98)
has an additional term which
exactly cancels the $1/\beta$ term which arises from application of
the Poisson summation formula.
With exponential accuracy we therefore have
\begin{equation}
\tilde \mu^2(\sigma,T) \> \simeq
\>  \frac{4}{\pi} \int_0^{\infty} dE \> \frac{1}{A(E)} \>
\frac{\sin^2 (E \sigma/ 2)}{E^2} \>.
\label{tilde amu2(sigma,T) Poisson}
\end{equation}
This is a big advantage as we do not have to expand the potential
term $ \> \ll S_1 \gg \> $ in Eq. (I.97) in inverse powers of
$\beta$. The only source of subasymptotic terms in
$ \> \ll S_1 \gg \> $ is then from the
$T$-integration from $ \> \sigma/2 \> $ to  $ \> \beta - \sigma/2 \> $
which simply gives a factor $ \beta - \sigma$.
Applying the Poisson formula to the kinetic term
$\tilde \Omega$ defined in Eq. (I.100) we obtain, again with
exponential accuracy
\begin{equation}
\tilde \Omega(\beta) \> = \> \frac{2}{\pi} \> \int_0^{\infty} dE \>
\left [ \> \ln A(E) \> + \> \frac{1}{A(E)} \> - \> 1 \>
\right ] \> + \frac{1}{\beta}
\> \left [ \> \ln A(0) \> + \> \frac{1}{A(0)} \> - \> 1 \>
\right ] \> .
\label{tilde Omega(beta) Poisson}
\end{equation}
We recall from (I) that the 2-point function may be written near
the pole as
\begin{equation}
G_2(p) \simeq \frac{1}{2} \> \int_0^{\infty} d\beta \>
\exp \left [ -\frac{\beta}{2} F(\beta,p^2) \right ] \>.
\label{2point func as beta int over F}
\end{equation}
Collecting all non-exponential terms the function $F(\beta,p^2)$
therefore has the large-$\beta$ expansion
\begin{equation}
F(\beta,p^2) \> \simeq \> F_0(p^2) \> + \> \frac{2}{\beta} \>
F_1(p^2) \> ,
\label{F(beta,p^2) large beta}
\end{equation}
where
\begin{equation}
F_0(p^2) \> = \> p^2 + M_0^2 - p^2 (1-\lambda)^2 + \Omega
- \frac{g^2}{4 \pi^2} \int_0^{\infty} d\sigma \> \int_0^{1} du \>
e\left( m \mu(\sigma), \frac{-i \lambda p \sigma}{\mu(\sigma)},u
\right )
\label{F0}
\end{equation}
is what we have used before on the nucleon pole
($p = i M_{\rm phys}$) and
\begin{equation}
F_1(p^2) =  \ln A(0) + \frac{1 - A(0)}{A(0)}
 + \frac{g^2}{8 \pi^2} \int_0^{\infty} d\sigma \>
\frac{\sigma}{\mu^2(\sigma)} \int_0^{1} du \>
e\left( m \mu(\sigma), \frac{-i \lambda p \sigma}
{\mu(\sigma)},u \right ) \> .
\label{F1}
\end{equation}
Note that the potential term in $F_0(p^2)$ develops a
small-$\sigma$ singularity which renormalizes the bare mass $M_0$
but $F_1(p^2)$ is finite.

Neglecting the exponentially suppressed terms and performing the
proper time integration we thus obtain the following
expression for the two-point function
\begin{equation}
G_2(p^2) \> \simeq \> \frac{ e^{-F_1(p^2)}}{F_0(p^2)} \> =
\>\exp \left[ \> - \ln F_0(p^2) - F_1(p^2) \> \right] \> .
\label{2point expressed by F0,F1}
\end{equation}
It is now very easy to calculate the residue $Z$ at the pole
(see Eq. (\ref{pole of 2point(p)}) )
by expanding
around the point $ \> p^2 = - M^2_{\rm phys} \> $ where $F_0$
vanishes. We obtain
\begin{equation}
Z \> = \> \frac{\exp\left[ -F_1(-M_{\rm phys}^2) \right ]}
{ F_0'(-M_{\rm phys}^2)}
\label{residue expressed by F0,F1}
\end{equation}
where the prime denotes differentiation with respect to $p^2$.
Explicitly we find
\begin{equation}
Z \> = \> \> \frac{ N_0 \> N_1}{D}
\label{residue explicit}
\end{equation}
where
\begin{eqnarray}
N_0 &=& \exp \left ( - \ln A(0) + 1 - \frac{1}{A(0)} \right )
\label{residue prefactor}\\
 N_1 &=& \exp\left[  \> - \frac{g^2}{8 \pi^2} \int_0^{\infty}
d\sigma \> \frac{\sigma}{\mu^2(\sigma)} \int_0^{1} du \>
e\left( m \mu(\sigma), \frac{\lambda \sigma M_{\rm phys}}
{\mu(\sigma)},u \right ) \right ]
\label{residue numerator} \\
\rm D &=& \> 1 - (1-\lambda)^2
 \> - \frac{g^2}{8 \pi^2} \> \lambda^2 \int_0^{\infty} d\sigma \>
\frac{\sigma^2}{\mu^4(\sigma)} \int_0^{1} du \> u \>
e\left( m \mu(\sigma), \frac{ \lambda \sigma M_{\rm phys}}
{\mu(\sigma)},u \right ) \nonumber \\
&=& \lambda \> .
\label{residue denominator}
\end{eqnarray}
In the last line the stationarity Eq. (\ref{var eq for lambda}) for
$\lambda$ was used to simplify the denominator $D$. Note that
this also applies to the case where one parametrizes
the profile function $A(E)$.
This demonstrates that
\begin{equation}
Z = \frac{ N_0 \> N_1}{\lambda}
\label{Z > 0}
\end{equation}
is always positive. It seems to be more difficult to prove in general
that $ Z \le 1 $  although all numerical calculations clearly
give this result.
Finally, it is again useful to check the variational result in
perturbation theory. With $A(0) = 1 + {\cal O}(g^2)$ one sees that
$N_0 = 1 + {\cal O}(g^4)$. Similarly
$ (1-\lambda)^2 = 1 + {\cal O}(g^4)$. Expanding $ N_1$ and $1/\lambda$
to order $g^2$ we obtain
\begin{eqnarray}
Z \> &=& \> 1 - \frac{g^2}{8 \pi^2} \> \int_0^{\infty} d\sigma \>
\int_0^{1} du \> (1 - u ) \>
\exp \left( - \frac{\sigma m^2}{2} \frac{1-u}{u} -
\frac{\sigma M^2_{\rm phys}}{2}u \right ) + {\cal O}(g^4) \nonumber \\
&=& \> 1 - \frac{g^2}{8 \pi^2} \> \int_0^{1} du \>
\frac{u (1 - u) } {M^2_{\rm phys} u^2 + m^2 (1 - u) } +
{\cal O}(g^4) \> .
\label{residue perturb check}
\end{eqnarray}
This coincides with what one obtains from the perturbative result
 for the self-energy (\ref{perturb self energy}) in the usual way.

Table~\ref{table: residue} contains the numerical values of the
residue obtained with
the different parametrizations as well as the perturbative result
from Eq. (\ref{residue perturb check})
\begin{equation}
Z_{ \rm perturb } \> = \> 1 \> - \> 0.38004 \> \alpha \> .
\label{Zpert}
\end{equation}
It is seen that for $\alpha$ near the critical value
appreciable deviations from the perturbative result
occur. For example, at $\alpha = 0.8$ perturbation theory says that
there is a probability of nearly $70 \%$ to find the bare particle
in the dressed one whereas the variational results estimate this
probability to be less than $50 \%$. It should be also noted that
the residue is {\it not} an infrared stable quantity, i.e. for
$m \to 0 $ $ \> Z \> $ also vanishes.
 From the variational equations one can deduce that
\begin{equation}
Z \> \buildrel m \to 0 \over \longrightarrow \> {\rm const.} \> \>
m^{\kappa}
\end{equation}
with $\kappa = \alpha/( \pi \lambda^2)$.
For massless mesons
the residue at the nucleon pole must vanish because
it is well known (e.g. from Quantum Electrodynamics) that in this case
the two-point function does not develop a pole but rather
a branchpoint at $p^2 = - M_{\rm phys}^2$.

\begin{table}
\begin{center}
\begin{tabular}{|c|c|c|c|c|} \hline
 ~~$\alpha$~~ & ~`Feynman'~ & ~`improved'~   & ~`variational'~
& ~perturbative~ \\ \hline
 0.1     & 0.96090         & 0.96087  & 0.96087  & 0.96200      \\
 0.2     & 0.91934         & 0.91914  & 0.91918  & 0.92399      \\
 0.3     & 0.87467         & 0.87418  & 0.87428  & 0.88599      \\
 0.4     & 0.82600         & 0.82494  & 0.82521  & 0.84798      \\
 0.5     & 0.77184         & 0.76996  & 0.77036  & 0.80998      \\
 0.6     & 0.70940         & 0.70610  & 0.70672  & 0.77198      \\
 0.7     & 0.63216         & 0.62597  & 0.62697  & 0.73397      \\
 0.8     & 0.51086         & 0.49187  & 0.49284  & 0.69597      \\
\hline
\end{tabular}
\end{center}
\caption{Residue at the pole of the two-point function for the different
parametrizations of the profile function. The heading `Feynman'
gives the result in the Feynman parametrization whereas
`improved' refers to the improved parametrization from Eq.
(\protect\ref{improved A(E)}). The residue calculated with the solution of
the variational equations is denoted by
`variational'. For comparison the perturbative result is also given.}
\label{table: residue}
\end{table}

\subsection{Variational equations for the off-mass-shell case}
\label{sec: var eq off mass shell}

It is also possible to apply the variational principle away from
the pole of the two-point function by varying Eq.
(\ref{2point expressed by F0,F1}). This gives
\begin{equation}
\delta \; F_0 (p^2)  \> + \> F_0 (p^2) \> \delta F_1 (p^2) \> = \> 0
\> .
\label{var eq off}
\end{equation}
Note that on mass-shell where $ F_0 $ vanishes the previous
variational equations follow. We will not elaborate on
Eq. (\ref{var eq off}) further
but only point out that the perturbative self-energy
(\ref{perturb self energy}) is {\em not} obtained from the off-shell
variational equations (\ref{var eq off}). In the limit
$\mu^2(\sigma) \to \sigma, A(E) \to 1, \lambda \to 1$ one rather finds
\begin{equation}
\Sigma_{\rm var}(p^2) \> \to \> - \frac{g^2}{4 \pi^2} \>
\ln \frac{\Lambda^2}{m^2} \> + \> \frac{g^2}{4 \pi^2} \> \int_0^1
du \> \ln \left [ 1 - \frac{p^2}{m^2} \frac{u^2}{1-u} \>
\right ]  + \frac{g^2}{4 \pi^2} \> \int_0^1 du \> u \>
\frac{p^2 + M_1^2}{m^2 (1-u) - p^2 u^2}
\end{equation}
which is an expansion of Eq. (\ref{perturb self energy})
around $ p^2 = - M_0^2$ (or $M_1^2$ which is the same in lowest order
perturbation theory). The reason for this somehow unexpected result
is the neglect of exponentially suppressed terms in deriving
Eq. (\ref{2point expressed by F0,F1}). Indeed it is easy to see
that one obtains the correct perturbative self-energy only if the
upper limit of the $\sigma$-integral
is kept at $\beta$ and not extended to infinity as we have done in
deriving Eq. (\ref{2point expressed by F0,F1}). The difference is
one of the many exponentially suppressed terms which we have
neglected. Thus, the off-shell variational equations
(\ref{var eq off}) only hold in the vicinity
of the nucleon pole and in order to investigate variationally the
two-point function far away (say near the meson production threshold
$p^2 = - ( M_{\rm phys} + m)^2$ ) one has to include consistently
all terms which are exponentially suppressed in $\beta$. This is
beyond the scope of the present work.

\section{Discussion and Summary}

\noindent
In the present work we have performed variational calculations
for the `Wick-Cutkosky polaron' following the approach which was
developed previously \cite{RoSchr}. We have determined
different parametrizations as well as the full variational solution
for the retardation function which enters
the trial action.
Since the
nucleon mass is fixed on the pole of the 2-point function the value
of the functional which we minimize is of no physical significance
but only a measure of the quality of the corresponding ansatz.
This is in contrast to the familiar quantum-mechanical case where an
upper limit to the ground-state energy of the system is obtained.
However, our calculation fixes the variational parameters with which
we then can calculate other observables of physical interest.

One of these quantities was the residue on the pole of the propagator
for which we have compared numerically the results of the variational
calculations to first order perturbation theory in Table~{table: residue}.
For small couplings all results for the residue agree,
since in this case
the variational approach  necessarily reduces to perturbation theory
independent of the value of the variational parameters.
What is rather remarkable is that for
larger couplings the three parametrizations of the profile function
in our variational approach
yield rather similar results, which are now of course different from
the perturbative calculation.  As
we have seen, the `improved' and `variational' actions have the
same singularity behaviour, for small relative times, as
the true action, so here one might expect some similarity in the
results.  This is however not true for the
`Feynman' parameterization which has a rather different form, so its
agreement with the other two is not
preordained.  This similarity is also exhibited in
Tables~\ref{table: var Feyn} and~\ref{table: var improved} for
$\lambda$ and $A(0)$, but of course not for the parameters $v$ and
$w$ which enter the respective profile functions and which are
`gauge' (i.e. reparametrization)-dependent quantities.
Also the critical coupling at
which real solutions ceased to exist was nearly identical in all
three parametrizations.
The similarity of the results for the different ans\" atze presumably
indicates that these results are not too
far away from the exact ones.

We were not only able to determine the critical coupling but also
to deduce qualitatively and quantitatively the  width which the
particle acquires beyond the critical coupling. This was achieved
by finding {\it complex} solutions of
the variational equations first approximately by an analytic approach
and then exactly by an iterative method which closely followed
the analytic procedure. Although the present
approach does not describe tunneling (which we expect to render the
system unstable even at small coupling constants but with
exponentially small width \cite{AfDel} ) the polaron variational
method is clearly superior to any perturbative treatment in this
respect.

We have concentrated mostly, although not exclusively, on the on-shell
2-point function, i.e. the nucleon propagator.
This corresponds to the limit where the proper time goes
towards infinity.  It is possible, however, to
go beyond the on-shell limit.  This was necessary, for example,
for the calculation of the residue of the
2-point function in Section \ref{sec: resi}.
Nevertheless, the residue is a quantity
which is calculated at the pole and thus only
requires off-shell information from an infinitesimal region around
it. This has the effect that the variational
parameters for the calculation of the residue are the same as the
on-shell ones.  As one moves a finite distance away
from the pole the variational parameters themselves become a
function of the off-shellness $p^2$
(see Section \ref{sec: var eq off mass shell}).

In conclusion, we think that the present variational approach
has yielded nonperturbative numerical results which look very
reasonable and are encouraging.
We therefore believe it worthwhile to try to extend it in several
ways. First, in a sequel to this work we will generalize
the present approach to the case with $n$ external mesons
and thereby study physical processes like meson production or meson
scattering from a nucleon.
This can be done by employing the quadratic trial function whose
parameters have been determined in the present work
on the pole of the 2-point function. Such a
`zeroth order' calculation is similar in spirit to a quantum
mechanical calculation in which
wave functions determined from minimizing the energy functional
are used to evaluate other observables. More demanding
is the consistent `first-order' variational calculation of
higher-order Green functions as this requires the amputation of
precisely the non-perturbative nucleon propagators which have been
determined in the present work. That this is indeed possible
will be demonstrated in another paper in this series.

Of course, finally we would like to apply these non-perturbative
techniques to theories which are of a more physical nature.
Among these one may mention scalar QED, the Walecka model
\cite{SeWa,Se} or QED. The latter two will require introduction of
Grassmann variables in order to deal with
spin in a path integral.  As such, this should not pose a
fundamental problem. A greater challenge, however, is to extend
such an approach beyond the quenched approximation or
to nonabelian theories where the light degrees of freedom
cannot be integrated out analytically.

\vspace{2cm}

\noindent
{\bf Note added}\\
 After completion of this work we became aware of the
pioneering work by K. Mano  \cite{Mano} in which
similar methods are applied to the Wick-Cutkosky model with zero
meson mass. Mano uses the proper
time formulation, the quenched approximation and the Feynman
parametrization for the retardation function to derive a
variational function for the self-energy of a scalar nucleon
(the expression following his Eq. (6.18))
which is identical with our Eq. (\ref{var inequality for Mphys})
after proper identification of quantities
is made. However, for minimizing the variational function Mano sets
(in our nomenclature) $ v = w (1 + \epsilon )$, expands to second
order in $\epsilon$  and finds an instability of the ground state for
$ g_{\rm Mano}^2/ 8 \pi M^2  > 0.34 $. Note that
$ g_{\rm Mano} = \sqrt{\pi} \> g$ so that this translates into a
critical coupling
$\alpha_c \approx 0.22$ which is much smaller than the value which
we obtain from the exact minimization.
In addition, in the present work we consider non-zero meson masses,
employ more general retardation functions, and calculate residue and
width of the dressed particle.

\vspace{3cm}
\noindent{\bf Acknowledgements}

\noindent
We would like to thank Dina Alexandrou and  Yang Lu for many helpful
discussions and Geert Jan van Oldenborgh for encouragement and
a careful reading of the manuscript.

\newpage

\noindent
{\Large\bf Appendix : An alternative expression for
$\Omega_{\rm var}$}

\renewcommand{\theequation}{A.\arabic{equation}}
\setcounter{equation}{0}

\vspace{0.5cm}

\noindent
Here we derive Eq. (\ref{Omega var expressed by g^2}) for the
kinetic term $\Omega$ when the variational equations are fulfilled.
We first perform an integration by parts
in the definition (\ref{Omega by A(E)}) of $\Omega$. The slow fall-off
of the variational profile function with $E$
\begin{eqnarray}
A(E) \>  \buildrel E \to \infty \over \longrightarrow \> &1& + \> \>
\frac{g^2}{4 \pi^2} \frac{1}{E^2} \int_0^{\infty} d\sigma \>
\frac{\sin^2(E \sigma/2)}{\sigma^2} \> + \> ... \nonumber \\
= \> &1& + \> \>
\frac{g^2}{16 \pi} \frac{1}{E} \> + \> ...
\label{var A(E) for large E}
\end{eqnarray}
leads to a contribution at $E = \infty$
\begin{equation}
\Omega_{\rm var} = \frac{g^2}{8 \pi^2} + \frac{2}{\pi}
\int_0^{\infty} dE \> \left [ - E \> \frac{A'(E)}{A(E)} +
\frac{1 - A(E)}{A(E)} \> \right ] \> .
\label{Omega with part int}
\end{equation}
We then write the variational equation (\ref{var eq for A(E)}) for
$A(E)$ in the form
\begin{equation}
\frac{1}{A(E)} - 1 \> = \> - \frac{g^2}{4 \pi^2} \> \int_0^{\infty}
d\sigma \> \frac{\sin^2(E\sigma/2)}{E^2 A(E)} \>
\frac{1}{\mu^4(\sigma)} \> X(\sigma)
\end{equation}
where
\begin{equation}
X(\sigma) \> = \> \int_0^1 du \> \left [ 1
+ \frac{m^2}{2}
\mu^2(\sigma) \frac{1-u}{u} -\frac{\lambda^2 M^2_{\rm phys} \sigma^2}
{2 \mu^2(\sigma)} u \right ] \> e \> \left ( m \mu(\sigma),
\frac{\lambda M_{\rm phys} \sigma}{ \mu(\sigma)}, u \right) \>.
\label{X(sigma)}
\end{equation}
The integration over $E$ can now be performed giving a factor
$\pi \mu^2(\sigma)/4 $ due to Eq. (\ref{amu2(sigma)}). Therefore we
have
\begin{equation}
\int_0^{\infty} dE \> \left [\frac{1}{A(E)} - 1 \>\right ] =
\> - \frac{g^2}{16 \pi} \>
\int_0^{\infty} d\sigma \> \frac{1}{\mu^2(\sigma)} \> X(\sigma)
\end{equation}
which is just one term in the expression (\ref{Omega with part int})
for $\Omega$. To get the other one we differentiate the variational
equation for $A(E)$ with respect to $E$ and observe that
\begin{equation}
\frac{\partial}{\partial E} \> \sin^2 \left( \frac{E \sigma}{2}
\right ) \> = \> \frac{\sigma}{E} \frac{\partial}{\partial \sigma }
\> \sin^2 \left( \frac{E \sigma}{2} \right ) \>.
\end{equation}
One has to be careful not to interchange the $E$-integration and the
$\sigma$-differentiation. We therefore perform an integration by parts
and obtain
\begin{eqnarray}
- \> \int_0^{\infty} dE \> E \> \frac{A'(E)}{A(E)} \> &=&
\frac{g^2}{4 \pi^2}
\int_0^{\infty} dE \> \frac{1}{E^2 A(E)} \> \Biggl [ \> \sigma
X(\sigma)
\frac{\sin^2(E \sigma/2)}{\mu^4(\sigma)} \> \Biggl |^{\infty}_0
\nonumber \\
&+&  \int_0^{\infty} d\sigma \> \frac{\sin^2(E \sigma/2)}
{\mu^4(\sigma)}
\>  \left (2 + \frac{\partial}{\partial \sigma} \sigma \> \right )
\> X(\sigma) \> \Biggr ]\nonumber \\
&=& \frac{g^2}{16 \pi} \left [ \> - \lim_{\sigma \to 0}
\frac{\sigma X(\sigma)}{\mu^2(\sigma)} +
\int_0^{\infty} d\sigma \> \frac{1}{\mu^2(\sigma)} \> \left (
2 + \frac{\partial}{\partial \sigma} \sigma \right ) \> X(\sigma)
\right ] \> .
\end{eqnarray}
Note that the boundary term at $\sigma = 0$ gives a contribution
because of $ X(0) = 1 $. This contribution exactly cancels the
term $g^2/ 8 \pi^2$ in Eq. (\ref{Omega with part int}).
Combining both terms  for $\Omega$
(which do not exist separately due to the slow
fall-off of $A(E)$ ) we obtain
\begin{eqnarray}
\int_0^{\infty} dE \> \left [ - E \frac{ A'(E)}{A(E)} \> + \>
\frac{1 - A(E)}{A(E)} \right ] &=& \frac{g^2}{16 \pi} \left [ - 1 +
\int_0^{\infty} d\sigma \> \frac{1}{\mu^2(\sigma)} \> \left (
\> 1 - \frac{\partial}{\partial \sigma}\sigma \> \right ) \>  \>
X(\sigma) \right ] \nonumber \\
&=&
\frac{g^2}{16 \pi} \left [ - 1 +
\int_0^{\infty} d\sigma \> X(\sigma) \left ( 1 + \sigma
\frac{\partial}{\partial \sigma} \> \right ) \frac{1}{\mu^2(\sigma)}
 \> \> \right ]
\end{eqnarray}
from which Eq. (\ref{Omega var expressed by g^2}) follows. In the
last line again an integration by parts has been performed but this
time there is no contribution from the boundary terms.

\end{document}